\documentclass[conference,a4paper]{IEEEtran}
\usepackage{amsmath,amssymb,amsfonts}
\usepackage{algorithmic}
\usepackage{graphicx}
\usepackage{textcomp}
\usepackage{xcolor}

\usepackage{siunitx}
\usepackage{hyperref}
\usepackage{amssymb}
\usepackage[capitalise,noabbrev]{cleveref}
\usepackage{xspace}
\usepackage{booktabs}
\usepackage[table]{xcolor}
\usepackage[normalem]{ulem}
\usepackage{authblk}

\usepackage[bibstyle=ieee,citestyle=numeric-comp]{biblatex}
\makeatletter
\DeclareRobustCommand\onedot{\futurelet\@let@token\@onedot}
\def\@onedot{\ifx\@let@token.\else.\null\fi\xspace}
\DeclareRobustCommand\nodot{\futurelet\@let@token\@nodot}
\def\@nodot{\ifx\@let@token.\else~\null\fi\xspace}

\newfont{\eaddfnt}{phvr8t at 12pt}

\def\cf{\emph{cf}\onedot}

\def\BibTeX{{\rm B\kern-.05em{\sc i\kern-.025em b}\kern-.08em
    T\kern-.1667em\lower.7ex\hbox{E}\kern-.125emX}}
\begin{document}

\title{
	ROI-GS: Interest-based Local Quality \\ 3D Gaussian Splatting\\
}

\author[1,2]{Quoc-Anh Bui}
\author[1]{Gilles Rougeron}
\author[2]{Géraldine Morin}
\author[2]{Simone Gasparini}
\affil[1]{Université Paris-Saclay, CEA, List, F-91120, Palaiseau, France}
\affil[2]{Université de Toulouse, Toulouse INP -- IRIT, France}
\renewcommand\Affilfont{\itshape\small}

\maketitle

\begin{abstract}
We tackle the challenge of efficiently reconstructing 3D scenes with high detail on objects of interest. 
Existing 3D Gaussian Splatting (3DGS) methods allocate resources uniformly across the scene, limiting fine detail to Regions Of Interest (ROIs) and leading to inflated model size. 
We propose ROI-GS, an object-aware framework that enhances local details through object-guided camera selection, targeted Object training, and seamless integration of high-fidelity object of interest reconstructions into the global scene. 
Our method prioritizes higher resolution details on chosen objects while maintaining real-time performance.
Experiments show that ROI-GS significantly improves local quality (up to \SI{2.96}{\decibel} PSNR), while reducing overall model size by \SI{\approx17}{\percent} of baseline and achieving faster training for a scene with a single object of interest, outperforming existing methods.
\end{abstract}


\section{Introduction}
Reconstructing complex 3D scenes with high-quality details for objects of interest remains a long-standing goal in computer vision and graphics. 
Like human vision, which naturally focuses on salient objects or depth-varying areas while leaving peripheral regions less resolved, this selective fidelity enables efficient perception. 
Similarly, computational methods benefit from concentrating resources on semantically or geometrically important regions. 
This capability is particularly vital in industrial applications and cultural heritage contexts.
Here, our work aims to create 3D digital twins of cultural heritage sites, facilitating preservation~\cite{Kong2023}, virtual exploration~\cite{Haibt2024}, and interactive museum experiences~\cite{Liu2024}, thus supporting archiving, education, and public participation while minimizing physical handling and degradation risks.

Recent advances in 3D Gaussian Splatting (\textbf{3DGS})~\cite{kerbl3Dgaussians} have rapidly gained traction as a powerful alternative for neural scene representation, offering explicit scene models, real-time rendering, and state-of-the-art visual quality. 
However, its explicit nature comes at the cost of a large memory footprint that can become prohibitive in large-scale scenes that require high detail in the internal contents of interest, as reported in ~\cite{kerbl3Dgaussians,niemeyer2025radsplat,kulhanek2025lodge}. This challenge is exacerbated by uniform scene reconstruction in most existing 3DGS methods, which allocates equal resources to all regions regardless of their importance. 
Prior work on level of detail (\textbf{LOD}) Gaussian Splatting ~\cite{Yu2023MipSplatting,kulhanek2025lodge,ren2024octree,cui2024LetsGo,seo2024flod,Milef2025Learning,yang2025lodgs,windisch2025lod,shen2025lod} or scene chunking ~\cite{hierarchicalgaussians24,10.1145/3687762} improves overall quality and scalability, but does not account for the presence of content of interest within the scene.
Efforts in object-centric~\cite{RoggeOC2DGS2025} and segmentation-aware~\cite{gaussian_grouping,lyu2024gaga,zhou2024drivinggaussian} modeling focus capacity on individual scene elements. 
While these methods enable precise object handling, they require coherent object masks across images and do not adaptively select or schedule views based on objects of interest.

\begin{figure}[!t]
\centering
\includegraphics[width=1\columnwidth]{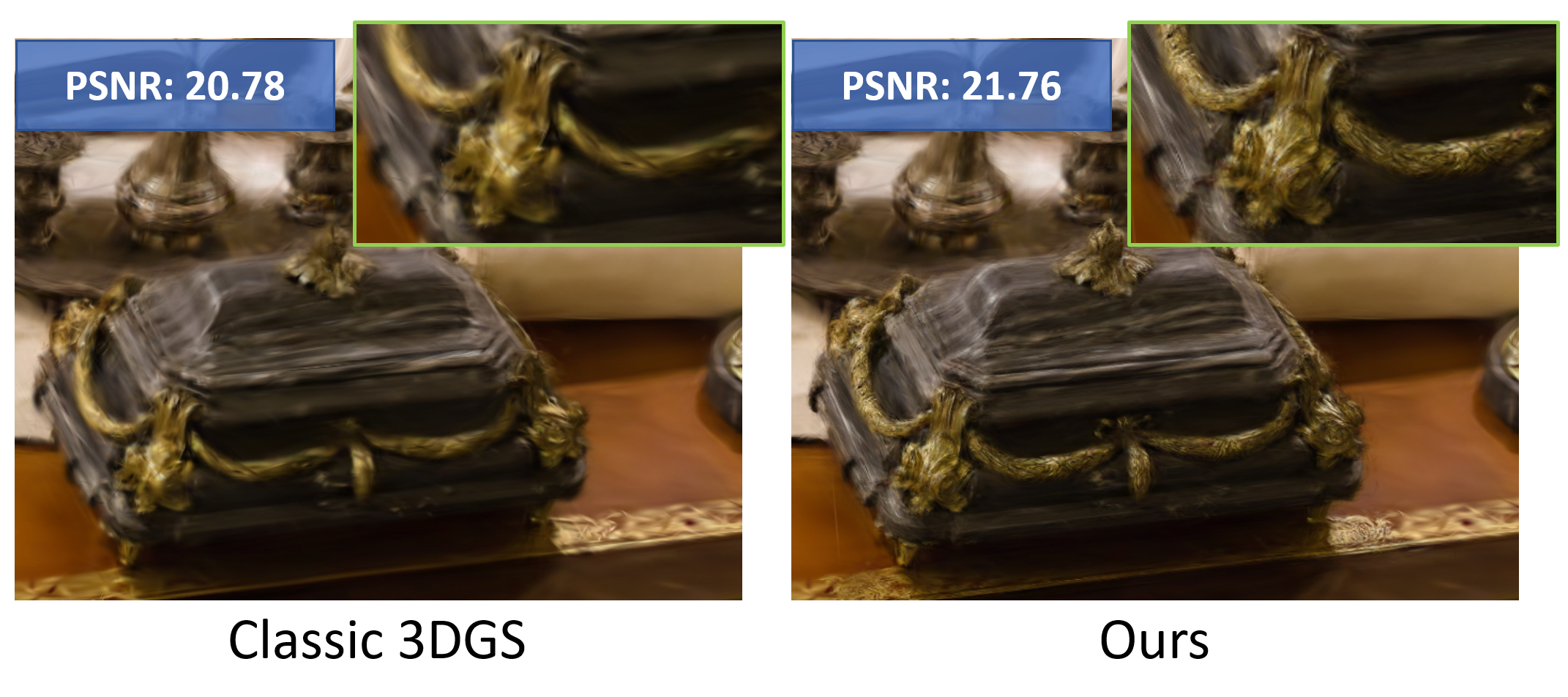}
\caption{
A comparison of our proposed method against the classic 3D Gaussian Splatting baseline shows that our method improved object detail with minimal increase in overall scene memory.
}
\label{fig:fig0_title}
\end{figure}

We address these gaps with \textbf{ROI-GS}, an object-aware 3DGS framework that reconstructs user-specified objects of interest using optimized image subsets and reintegrates them into the overall scene model. 
Our pipeline includes a scene decomposition module based on object-specific automatic camera selection, enabling independent training of scene components.
By focusing computational resources and details on user-selected ROIs, the approach achieves fine-grained ROI optimization without significantly increasing the overall memory footprint.
Leveraging the explicit nature of 3DGS representation, we seamlessly replace lower-LOD Gaussians in the base scene with refined ROI Gaussians, producing a unified model that combines global context at moderate quality with high-quality local regions, all rendered in real-time. 

\begin{figure*}[!t]
\centering
\includegraphics[width=.83\linewidth]{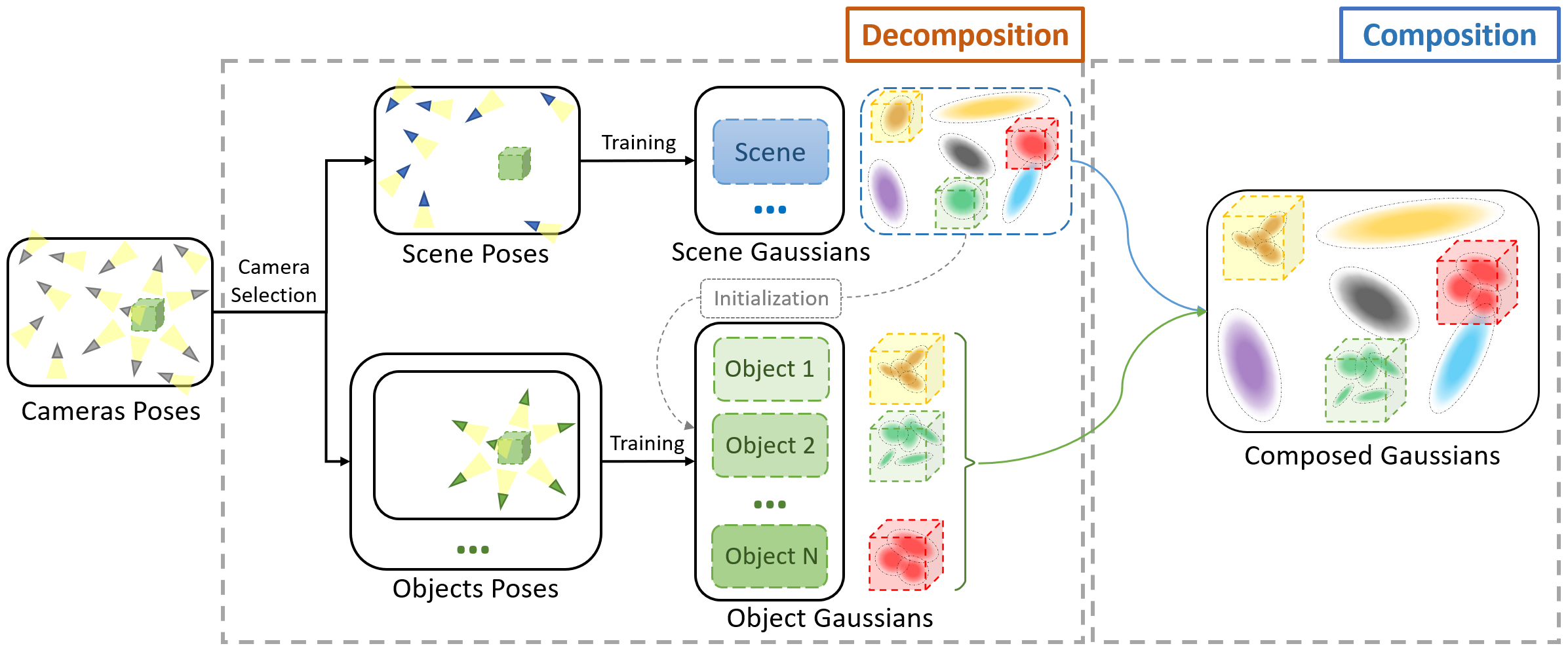}
\caption{
An overview of the proposed Region of Interest-Focused Gaussian Splatting (ROI-GS) framework, consisting of two stages: scene decomposition and composition.
In decomposition, the scene is divided into Scene and Objects groups, with camera sets automatically selected for each GS training.
The trained Scene-GS model is used to initialize the Object-GS models.
The composition stage integrates high-detail Object-GSs and the global Scene-GS to produce high-quality real-time renderings with enhanced detail for objects in the ROIs.
}
\label{fig:pipeline}
\end{figure*}

The main contributions of this work are: (\textit{i}) we introduce region-aware reconstruction into the 3DGS paradigm through object-guided view selection and  targeted Object training; 
(\textit{ii}) we improve reconstruction quality using an advanced image selection strategy based on model optimization that maximizes ROI coverage and occupancy; 
and (\textit{iii}) we propose a simple yet effective composition strategy that integrates high-fidelity Object Gaussians into the overall scene model. 
Together, these contributions enable ROI-GS to enhance object-level detail in complex scenes without sacrificing real-time rendering performance under a practical number of ROIs. 

\section{Method}
\label{sect:method}

Our ROI-GS pipeline builds on the standard 3DGS framework, which models the scene as a set of anisotropic 3D Gaussians.
It takes as input a set of 2D color images, along with camera poses, intrinsic parameters, and a sparse point cloud, 
producing highly realistic real-time renderings with enhanced LOD in regions of interest (\cf ~\cref{fig:pipeline}).

\subsection{Object-Focused Camera Selection.}
\label{sect:method:select}
We propose a two-step ROI-focused camera selection strategy. First, each user-specified ROI is defined by an axis-aligned bounding box (AABB) enclosing the object of interest.
In the initial filtering step, we select all views that observe the object by identifying cameras that capture at least one 3D keypoint within the bounding box, based on point-to-image visibility from SfM data. 
This first filtering significantly narrows down the set of candidate images. 
However, using all these views to train an Object-GS may degrade object detail due to low resolution from distant cameras and also reduce computational efficiency. 
To address this, we introduce a refinement step that assesses and selects the most informative and optimal views for high-quality object reconstruction. 
We initially used a simple approach based on static criteria, such as the distance between the ROI and viewpoints, the projected area of the ROI’s bounding box on 2D images, or the number of reconstructed keypoints within the box and visible to each camera. 
While effective in identifying views rich in ROI information, these criteria could introduce spatial bias, favoring regions with dense camera clusters and resulting in incomplete or unbalanced coverage from diverse perspectives. 
To mitigate this, we adopt an \textbf{advanced selection} strategy based on model-driven optimization.
Prior works on View selection in GS~\cite{li2024frequency,Polyzos2025ActiveInitSplat,strong2025bestsense,li2025activesplat} reduce image redundancy and maximize scene coverage by selecting globally informative views, but they do not account for object-specific camera relevance.
Our advanced selection approach relies on ActiveInitSplat~\cite{Polyzos2025ActiveInitSplat}, with adaptations for object-focused optimization. 
The views are ordered based on the ROI's 3D point quality score: the next view is selected as the one that maximizes the improvement of this score.
Specifically, this refinement selection module considers two terms of the sparse point distribution: the point density and voxel occupancy within the ROI bounding box to identify any sparsely covered areas or angles.
A Gaussian Process-based (GP) model is used to guide this selection.
Furthermore, by incorporating the static criteria mentioned above as parameters, we achieve more accurate and consistent camera selection for the target object, leading to better-aligned Gaussian optimization and ultimately higher-fidelity ROI reconstructions. 
The original model uses $6$ input viewpoint parameters ($3$ for position and $3$ for orientation). 
Our extended version adds $3$ static parameters (distance, projected AABB area, and number of keypoints), totaling $9$ inputs. 
We compare the $6$-params and $9$-params versions in \cref{sect:exp:abla}.

\subsection{GS Training}
\label{sect:method:train}

Once the relevant images are selected and grouped, we train separate high-quality Gaussian Splatting models for each object (Object-GS) and one for the global scene (Scene-GS).
The Scene-GS model is first trained for $20K$ iterations using all scene images, excluding the subset of close-up views reserved for ROI training. To maintain consistency, a portion of the ROI images can optionally be retained in the Scene-GS training to give the global model a coarse understanding of the object’s geometry.
In our experiments, \SI{50}{\percent} of the ROI selected images are included in Scene-GS training.

In early experiments, we attempted to initialize Object-GS from the sparse SfM point cloud. 
However, neglecting the global representation of the trained Scene-GS model proved inefficient. 
The next implementation idea was to initialize only Scene-GS Gaussians restricted to the ROI’s volume, and continue optimizing only within this region.
However, the lack of out-of-the-box context led to floating artifacts, as the model hallucinated missing geometry beyond the object of interest, significantly degrading reconstruction quality. 
To overcome this, we initialize each Object-GS directly from the full Scene-GS model.
This approach provides both contextual grounding and consistently yields higher-quality object reconstructions, making it a more effective and efficient starting point for ROI refinement.
Each Object-GS is then trained independently using only the subset of selected images focused on the target object. 
The object training follows standard 3DGS optimization for $30K$ iterations to ensure sufficient reconstruction quality. 
Optimization is applied to the entire image, updating Gaussians even in the vicinity of the object box and the background, to prevent floating artifacts within the ROI. 
However, the Gaussian densification process is confined to the ROI volume for $15K$ iterations, keeping detail enhancement limited within the target region. 
Additionally, a pruning strategy is applied during training to remove non-contributing Gaussians, many of which were inherited from the full Scene-GS initialisation, thereby reducing the memory footprint of the Object-GS model. 
Focusing on a bounded region with high-resolution inputs allows for significantly finer LOD in this region than the global model. 
Object-GS can thus capture high-fidelity representations, including geometric details and texture appearance that could otherwise be blurred or undersampled in a uniformly trained entire-scene model.

\subsection{Scene–Objects Composition with Gaussians.}
\label{sect:method:compo}
For the inference step, we integrate the Object-GS models back into the global Scene-GS. 
This composition is simple and efficient thanks to the explicit nature of 3DGS scene representation. 
For each object, we replace the original Scene-GS Gaussians that lie inside the object's bounding box with the corresponding fine details Object-GS ones. 
Since all Gaussians share a common coordinate system from SfM, the replacement is seamless and co-registered. 
The result is a combined Gaussian set representing the entire scene, where each ROI is now filled with its high-quality Gaussians.

\section{Experiments}
\label{sect:exp}

\begin{figure*}[t!]
	\centering
	\includegraphics[width=\linewidth]{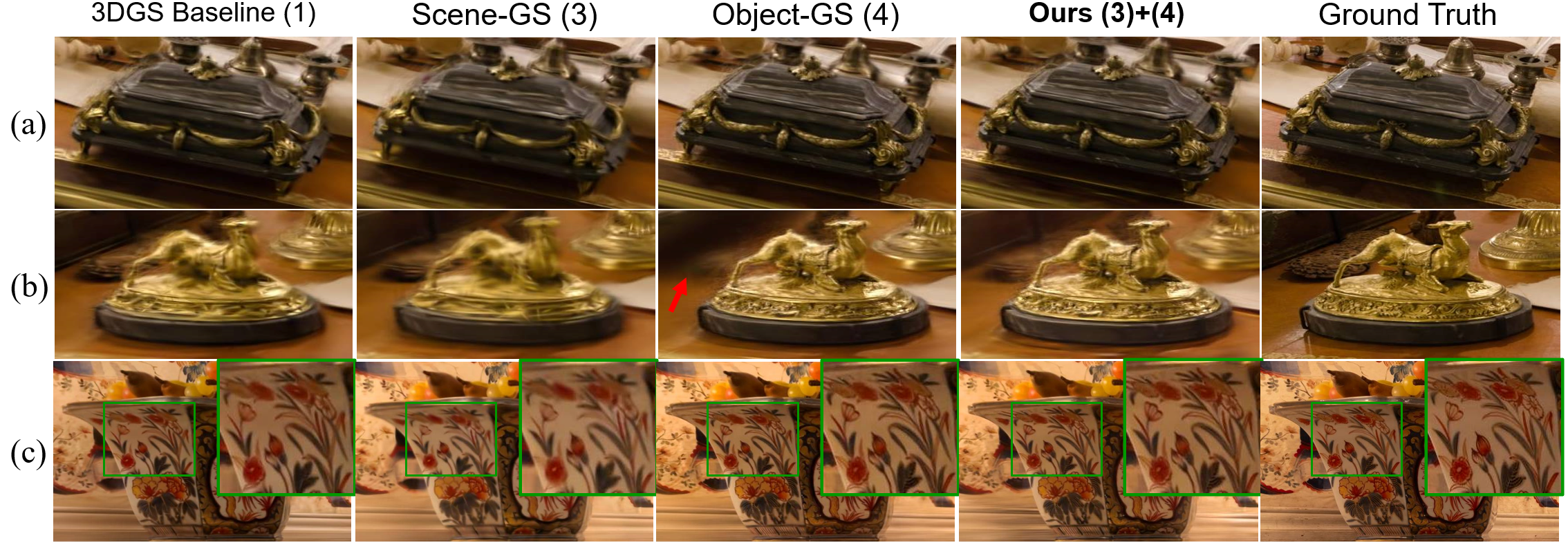}
	\caption[]{A visual comparison of our method and the baseline. Blurred area are indicated by red arrows.}
	\label{fig:compa}
\end{figure*}

In this Section, we discuss the implementation details of our ROI-GS method and present its performance in enhancing object-level reconstruction in a complex real-world scene.

\subsection{Experimental setup.}
\label{sect:exp:setup}

We conduct experiments on two scenes, a \textit{Bureau} and a \textit{Dining room}, captured at the 18th-century cultural heritage site, the \textit{Hôtel de la Marine} in Paris~\cite{hotelwebsite}.
The \textit{Bureau} capture consists of $761$ images at $8192\times5464$ pixels, covering the working room, including objects on a desk, such as a coffret and statuettes.
The \textit{Dining Room} capture includes $916$ images at the same resolution, plus $222$ close-up $3024\times4032$ images, featuring interesting objects like ornate bowls, vases, and chairs.
All images were downsampled by a factor of 4 to reduce storage and computation costs.
Our implementation is based on the 3DGS framework~\cite{kerbl3Dgaussians}. All trainings are performed using a single \SI{95}{\giga\byte} H100 GPU on a DGX workstation.

For fair comparison and objective performance evaluation, we consider a single object of interest in \cref{tab:quanti_eval} and \cref{tab:ablation}.
Nevertheless, the method is inherently scalable: \cref{tab:quanti_multi} experiments on multiple objects, and different objects are handled independently.
Out of $335$ images with at least one visible keypoint in the coffret AABB, $21$ are used as a test set to evaluate the model.
The baseline models used for comparison are: (1) Full Scene (3DGS), which is trained on all images in the scene, and (2) Full Object, which is trained on all images where the object of interest is visible.
Both models are optimized for 50K iterations to ensure maximizing the scene reconstruction quality.

We evaluated the Object-Focused Camera Selection technique on two versions of the Advanced GP-based model.
In our experiments, for ROI training we select $150$ from $314$ non-test ROI-visible images.
Selecting more images slightly degraded reconstruction quality, as distant or less informative views reduced effective resolution.
Both the original GP-based model (Advanced 6-params GP) and its extended version (Advanced 9-params GP), described in \cref{sect:method:select}, select the first $150$ cameras while preserving their order.
Unlike the default 3DGS pipeline, where camera order is irrelevant, the advanced selection strategy defines an order for selected views. 
During training, we retain this order instead of applying random shuffling, which helps the GS model converge faster and achieve higher reconstruction quality as shown in \cref{tab:ablation}.

\subsection{Quantitative and qualitative evaluation.}
\label{sect:exp:eval}

We use standard metrics such as PSNR and SSIM for consistent and meaningful comparisons. Since our concern is the quality improvement of the object representation, evaluating the entire image would be less relevant.
Therefore, scores are computed only on the pixels within the projected AABB.

\begin{table}[t!] 
    \centering
    \caption{Evaluation of our method compared to baselines on a ROI. 
    }
    \label{tab:quanti_eval}
    \resizebox{0.5\textwidth}{!}{
        \begin{tabular}{l|*{5}{c}}
        Methods$\vert$Metrics
        & {PSNR\textsuperscript{$\uparrow$}} & {SSIM\textsuperscript{$\uparrow$}} & {\#G\textsuperscript{$\downarrow$}} & {\#G in box} & {Train} \\

        \hline
        \textit{\underline{Baselines}: 50K iters} \\
        Full Scene images - 3DGS (1)   & {21.23} & {0.917} & {3.72M} & {8.73K} & {98m} \\
        Full Object images (2)     & {21.57} & {0.922} & {3.28M} & {16.12K} & {94m} \\

        \hline
        \textit{\underline{Ours}} \\
        \textcolor{gray}{Scene-GS: 20K iters} (3)   & \textcolor{gray}{20.78} & \textcolor{gray}{0.911} & \textcolor{gray}{3.02M} & \textcolor{gray}{6.99K} & \textcolor{gray}{35m} \\
        \textcolor{gray}{Object-GS: 30K iters} (4)     & \fcolorbox{red}{white}{\textcolor{gray}{21.99}} & \textcolor{gray}{0.924} & \textcolor{gray}{1.04M} & \textcolor{gray}{74.74K} & \textcolor{gray}{35m} \\

        \textbf{Composition (3)+(4)}     & \cellcolor{orange!50}{21.76} & \cellcolor{orange!50}{0.924} & \cellcolor{orange!50}{3.09M} & {74.74K} & \cellcolor{orange!50}{70m} \\

        \end{tabular}
    }
\end{table}
\begin{table}[h] 
    \centering
    \caption{Evaluation of our method on different objects of 2 scenes. 
    }
    \label{tab:quanti_multi}
    \resizebox{0.5\textwidth}{!}{
        \begin{tabular}{l|*{4}{c}|*{3}{c}}
        Dataset Scene & \multicolumn{4}{c|}{Bureau} & \multicolumn{3}{c}{Dining room} \\ 
        {PSNR\textsuperscript{$\uparrow$}}
        & {Coffret} & {Statue} & {Chair1} & {Sofa} & 
        {Vase} & {Bowl} & {Chair2} \\

        \hline
        3DGS baseline (1)           & {21.23} & {21.88} & {23.55} & {23.94} &
                                        {23.65} & {25.50} & {26.90} \\

        Our composition           & \cellcolor{orange!50}{21.76} & \cellcolor{orange!50}{22.31} & \cellcolor{orange!50}{24.01} & \cellcolor{orange!50}{24.19} &
                                        \cellcolor{orange!50}{24.81} & \cellcolor{orange!50}{26.30} & \cellcolor{orange!50}{27.50} \\

        \hline
        Peak Improvement           & {0.93} & {0.93} & {1.10} & {0.80} &
                                        \fcolorbox{red}{white}{2.96} & {2.13} & {2.63} \\



        \end{tabular}
    }
\end{table}

\cref{tab:quanti_eval} presents quantitative comparisons between our method and baselines on a single ROI. 
The proposed composition approach, which combines Scene-GS (3) and Object-GS (4), achieves a higher ROI score than all baselines. 
The table also lists the total number of Gaussians (\#G) and those within the ROI box.
\cref{tab:quanti_multi} reports PSNR scores for different objects across the two scenes, highlighting the superiority, flexibility, and scalability of our method.
Quality clearly improves across all objects, with a peak improvement per test image up to \SI{2.96}{\decibel} for the ROI Vase in the \textit{Dining Room}.

\cref{fig:compa} presents qualitative evaluations on three different ROIs.
The proposed method yields noticeably sharper object details than both the baseline and Scene-GS. 
Rows (a) and (b) illustrate the precise reconstruction of complex geometry , while row (c) captures the vase’s intricate surface texture. 
In Object-GS (4), optimization is applied not only to Gaussians inside the box but also to outer ones, achieving the highest PSNR within the AABB (red box in \cref{tab:quanti_eval}). 
This highlights the strength of the ROI reconstruction in capturing objects fine details but limiting scene coverage, leading to reduced overall scene quality (red arrow in \cref{fig:compa}b).

\subsection{Ablations.}
\label{sect:exp:abla}

\begin{table}[h] 
    \centering
    \caption{PSNR, SSIM scores for ablations. 
    }
    \label{tab:ablation}
    \resizebox{0.5\textwidth}{!}{
        \begin{tabular}{l|*{5}{c}}
        Methods$\vert$Metrics
        & {PSNR\textsuperscript{$\uparrow$}} & {SSIM\textsuperscript{$\uparrow$}} \\

        \hline
        \textbf{Ours}     & \cellcolor{orange!50}{21.76} & \cellcolor{orange!50}{0.924} \\
        \hline
        
        No Scene-GS Init + 9-params GP (1)          & {21.65} & {0.919} \\
        No 3 static ROI params (6-params GP) (2)    & {21.20} & {0.908} \\
        No Advanced GP-based model (3)              & {19.18} & {0.877} \\
        
        \end{tabular}
    }
\end{table}

We evaluate the impact of Advanced Camera Selection and Scene-GS initialization for object training in \cref{tab:ablation}. 
In ablation setting (1): we use SfM point cloud initialization instead of trained Gaussians from Scene-GS, combined with Advanced 9-parameters GP camera selection. 
This results in a slight drop in ROI reconstruction quality, compared to the full method. 
Settings (2) and (3) both use Scene-GS initialization. (2) employs the original GP-based model without the three static ROI information, while (3) uses the simple selection based on static criteria only instead of the Advanced GP-based selection. 
The results highlight the effectiveness of the Advanced Selection approaches using GP model-based optimization. 
Moreover, incorporating additional object-specific information further improves the GP model.

\section{Conclusion and Perspectives}
\label{sect:conclusion}

We presented ROI-GS, a 3DGS method that enhances the level of detail for objects of interest by leveraging camera selection and a simple composition strategy.
Our approach highlights its flexibility in enhancing object fidelity when applied to any global optimization model without compromising surrounding scene quality.
While ROI-GS enables efficient and high-quality object view synthesis, there are limitations.

First, our method relies on bounding boxes for object delineation, which can yield imprecise boundaries. 
Incorporating object-level segmentation could improve spatial accuracy and reduce visual artifacts in cluttered or occluded areas.

Second, LOD generation during GS training is currently static at render time. 
Introducing dynamic LOD adjustment based on the camera viewpoint could enable smarter resource allocation and better performance.
Additionally, ROI-GS training time scales linearly with the number of objects, and model size depends on object count and size. 
An excessive number of Gaussians could negatively impact real-time rendering performance, making dynamic LOD essential. 
For example, distant objects could simply ignore the Object-GS and fall back to Scene-GS.

Finally, the composition strategy may produce minor inconsistencies near ROI boundaries, as center-inside Gaussians may extend beyond the AABB. 
Although generally negligible in practice, improved spatial filtering or blending strategies could further enhance compositional accuracy.

\hfill

{\small
\noindent\textbf{\uppercase{Acknowledgments.}} 
We would like to thank the \textit{Centre des Monuments Nationaux} and the \textit{Hôtel de la Marine} team for giving us access to this magnificent place.
}

{\AtNextBibliography{\small}
\printbibliography
}

\end{document}